\documentclass[journal]{IEEEtran}

\usepackage{cite}
\usepackage{amsmath,amssymb,amsfonts}
\usepackage{bm}
\usepackage{algorithmic}
\usepackage{graphicx}
\usepackage{textcomp}
\usepackage{xcolor}
\usepackage{booktabs}
\usepackage{array}
\usepackage{url}
\usepackage[hidelinks]{hyperref}

\newcommand{\vect}[1]{\bm{#1}}          
\newcommand{\mat}[1]{\bm{#1}}           
\newcommand{\tp}{\mathsf{T}}             
\newcommand{\herm}{\mathsf{H}}             
\newcommand{\E}{\mathbb{E}}             
\newcommand{\Cset}{\mathbb{C}}             
\newcommand{\tr}{\operatorname{tr}}     
\newcommand{\rank}{\operatorname{rank}}
\newcommand{\diag}{\operatorname{diag}}
\newcommand{\vecop}{\operatorname{vec}}
\newcommand{\Real}{\operatorname{Re}}
\newcommand{\Imag}{\operatorname{Im}}
\newcommand{\CN}{\mathcal{CN}}          
\newcommand{\CRB}{\mathrm{CRB}}
\newcommand{\FIM}{\vect{J}}
\newcommand{\defeq}{\triangleq}

\begin{document}
\bstctlcite{IEEEexample:BSTcontrol}

\title{Integrated Sensing and Communications: \\ A Tutorial
on Mathematical Foundations, \\ Signal Processing and System Design for 6G}

\author{Sabit~Ekin,~\IEEEmembership{Senior~Member,~IEEE}
        \thanks{This tutorial is prepared as a self-contained pedagogical introduction to
        integrated sensing and communications. It is intended for graduate students and
        practitioners entering the field. 

Sabit Ekin is with the Department
of Engineering Technology and Industrial Distribution, Texas~A\&M~University, College Station, TX 77843
USA (e-mail: sabitekin@tamu.edu). 


}
}

\markboth{Tutorial on Integrated Sensing and Communications}%
{Ekin: A Tutorial on Integrated Sensing and Communications}

\maketitle

\begin{abstract}
Integrated sensing and communications (ISAC) is a cornerstone of sixth-generation (6G)
wireless networks: a single waveform, aperture, and hardware platform \emph{simultaneously}
conveys information and senses the physical environment. Its literature, however, is spread
across radar signal processing, estimation and information theory, and wireless
communications, which makes it difficult for a newcomer to assemble the full mathematical
picture. This tutorial fills that gap. Starting from complex-baseband signal models, we build
the required foundations step by step, with self-contained derivations. We develop the sensing
toolkit (matched filter, ambiguity function, maximum-likelihood estimation, Fisher information,
the Cram\'er--Rao bound (CRB), and Neyman--Pearson detection) and the communication toolkit
(Shannon and MIMO capacity, water-filling, and multiuser precoding), and then unify them into
the fundamental limits of ISAC: the CRB--rate region, the capacity--distortion function, and
the deterministic--random tradeoff. On these foundations we treat waveform design (OFDM radar
with two-dimensional FFT processing, OTFS, and jointly optimized waveforms), transmit
beamforming as explicit optimization, including semidefinite relaxation, and receiver-side
estimation and detection. We then survey the architectures that scale ISAC to the network
level (millimeter-wave and terahertz operation, reconfigurable intelligent surfaces, and
cell-free networks), together with applications, standardization, and security. A worked
MIMO-OFDM design example with concrete numerology ties the mathematics together. Throughout,
notation is unified and tabulated, key results are derived rather than merely cited, and
seminal contributions are placed in context, so that the reader finishes with both the
mathematical fluency and the literature map required to begin research in ISAC.
\end{abstract}


\IEEEpeerreviewmaketitle

\section{Introduction}
\IEEEPARstart{F}{or} most of the twentieth century, radar and communication systems were
designed, built, and studied as separate disciplines~\cite{liu2023seventy}. Radar engineers
sought to estimate the parameters of physical targets (their range, velocity, and
angle), while communication engineers sought to convey information reliably and at high rate.
The two communities developed largely disjoint mathematical toolkits: estimation and
detection theory on one side, information theory and coding on the other, meeting mainly through the lens of spectral coexistence~\cite{zheng2019radar}. Integrated sensing
and communications (ISAC) dissolves this boundary. In an ISAC system, a single transmitted
waveform, radiated from a shared antenna aperture and processed by a shared hardware chain,
performs \emph{both} functions at once~\cite{liu2022dual,liu2020joint}.

\begin{figure}[!t]
\centering
\includegraphics[width=0.99\columnwidth, trim={1.4cm 0.5cm 1.5cm 0.6cm},clip]{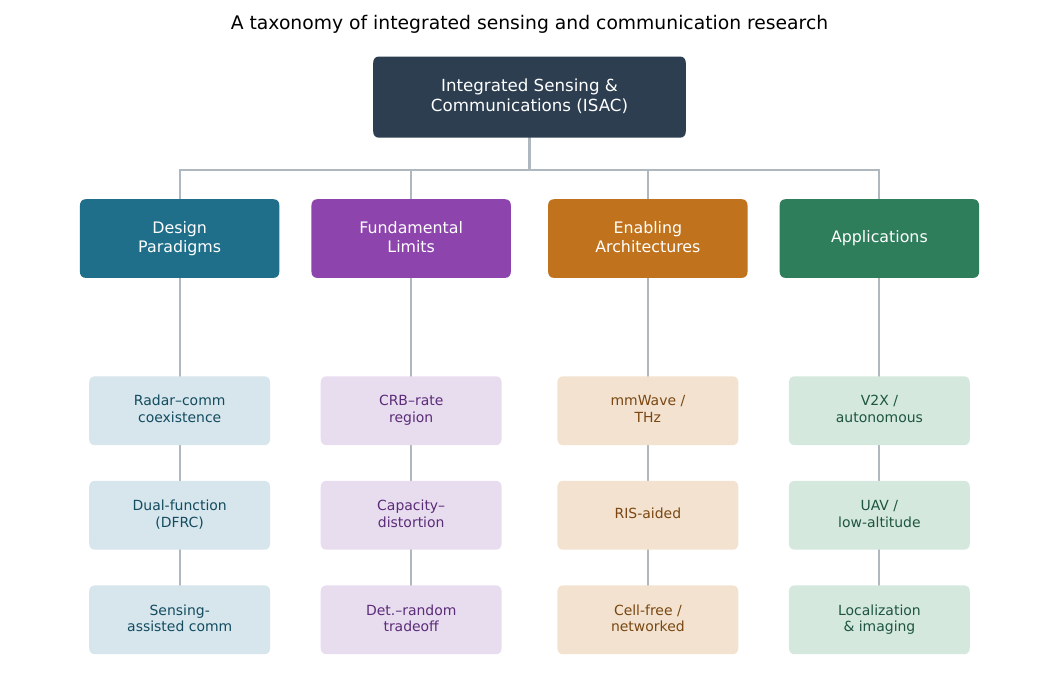}
\caption{A taxonomy of ISAC research along four axes: design paradigms, fundamental limits,
enabling architectures, and applications. In the design-paradigm column, moving from
radar--communication coexistence to dual-function (DFRC) design increases the degree of resource
sharing and the potential integration gain, at the cost of tighter coupling between the two
functions; the fully joint regime discussed in the text continues this progression. This
tutorial focuses on the dual-function and joint-design regimes.}
\label{fig:taxonomy}
\end{figure}

The promise of ISAC is compelling: improved spectral, hardware, and energy efficiency, and
the emergence of a \emph{perceptive network} that senses its surroundings as a native
service~\cite{zhang2021perceptive}. The field has consequently grown explosively, to the
point where Recommendation ITU-R M.2160 lists integrated sensing and communication among the
six usage scenarios of IMT-2030, the 6G framework~\cite{itu2023m2160}, and industry regards it
as a defining capability of sixth-generation (6G) wireless~\cite{liu2022dual,wild2023from}. Fig.~\ref{fig:taxonomy} organizes the research landscape into design paradigms,
fundamental limits, enabling architectures, and applications. Within the design paradigms,
integration admits degrees: at one extreme, radar and communication systems merely
\emph{coexist}, sharing spectrum while remaining separate; a \emph{dual-function} design reuses one
hardware platform and waveform for both tasks; and a fully \emph{joint} design optimizes the
transmitted signal from first principles against a combined objective. A related paradigm,
sensing-assisted communication, uses sensing outputs such as user positions to aid beam
management and link adaptation. This tutorial concentrates on
the dual-function and joint regimes, where the mathematics of the two disciplines genuinely merge.

The breadth of the field is, however, an obstacle for the
newcomer: to understand ISAC one must simultaneously be fluent in radar signal processing,
in estimation and information theory, and in multi-antenna communications, and the relevant
results are scattered across three literatures that rarely appear together.

\subsection{Purpose and Philosophy of This Tutorial}
This tutorial is written for the graduate student or practitioner who is beginning research
in ISAC and who wants a single, self-contained mathematical foundation. Our guiding
principles are:
\begin{itemize}
\item \emph{Derive, don't just cite.} Every central result (the ambiguity function, the
Cram\'er--Rao bound, the MIMO capacity, the CRB--rate region) is developed from first
principles, with enough intermediate steps that the reader can reproduce it.
\item \emph{One notation, everywhere.} Sensing and communications use overlapping symbols for
different quantities. We fix a single, unified notation (Table~\ref{tab:notation}) and use it
consistently across both functionalities.
\item \emph{Build in layers.} We first establish the shared signal model, then the sensing
toolkit, then the communication toolkit, and only then combine them into the ISAC-specific
limits and designs. Each layer depends only on the ones before it.
\item \emph{Connect math to the literature.} Seminal contributions are cited at the exact
point where their result is derived, so the reader simultaneously builds mathematical fluency
and a map of the field.
\end{itemize}

\subsection{Prerequisites}
We assume familiarity with linear algebra (vectors, matrices, eigendecomposition), complex
analysis at the level of Fourier transforms, and probability (random vectors, Gaussian
distributions, expectation). Prior exposure to digital communications and to the basics of
detection and estimation is helpful but not required; the necessary results are derived or
stated as we go. Readers already versed in one of the two disciplines may treat the
corresponding foundational section (Section~\ref{sec:radar} or~\ref{sec:comm}) as a review.

\subsection{Learning Objectives}
After working through this tutorial, the reader should be able to
(i) write down the complex-baseband signal model for a MIMO ISAC system serving communication
users while sensing targets;
(ii) derive the Cram\'er--Rao bound on target-parameter estimation and interpret the
ambiguity function of a waveform;
(iii) compute the MIMO channel capacity and multiuser achievable rates;
(iv) explain and derive the fundamental sensing--communication tradeoffs (CRB--rate,
capacity--distortion, deterministic--random);
(v) formulate waveform-design and beamforming problems as explicit optimization programs; and
(vi) place the major architectures, applications, and open problems of ISAC within this
mathematical framework.

\subsection{Organization and Reading Guide}
\begin{figure}[!t]
\centering
\includegraphics[width=0.99\columnwidth]{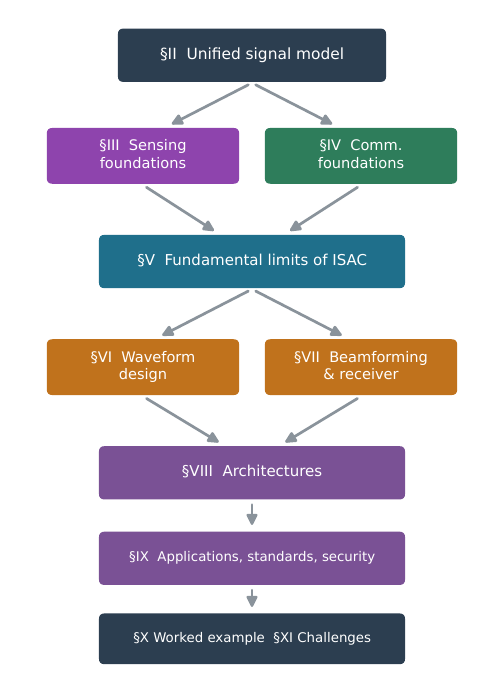}
\caption{Dependency structure of this tutorial. Arrows denote prerequisite relationships: the
unified signal model underpins the sensing and communication foundations, which combine into
the fundamental limits of ISAC; these in turn support the waveform-design and beamforming
methodologies and, ultimately, the architectures and applications.}
\label{fig:roadmap}
\end{figure}

Fig.~\ref{fig:roadmap} shows the dependency structure of the tutorial. Section~\ref{sec:model}
establishes the unified signal model. Sections~\ref{sec:radar} and~\ref{sec:comm} develop the
sensing and communication foundations, respectively. Section~\ref{sec:limits} unifies them
into the fundamental limits of ISAC. Sections~\ref{sec:waveform} and~\ref{sec:beamforming}
treat waveform design and beamforming/receiver processing as design problems built on those
limits. Section~\ref{sec:arch} surveys enabling architectures, and
Section~\ref{sec:apps} the applications, standardization, and security landscape.
Section~\ref{sec:example} presents a fully worked MIMO-OFDM design example, and
Section~\ref{sec:challenges} closes with open challenges. A reader wanting only the essential
path may follow Sections~\ref{sec:model}--\ref{sec:beamforming} in order.

\begin{table}[!t]
\centering
\caption{Summary of Principal Notation}
\label{tab:notation}
\renewcommand{\arraystretch}{1.18}
\begin{tabular}{p{1.9cm} p{6.0cm}}
\toprule
\textbf{Symbol} & \textbf{Meaning} \\
\midrule
\multicolumn{2}{l}{\emph{Operators and conventions}}\\
$a,\ \vect{a},\ \mat{A}$ & scalar, column vector, matrix \\
$(\cdot)^{\tp},(\cdot)^{*},(\cdot)^{\herm}$ & transpose, conjugate, conjugate transpose \\
$\tr(\cdot),\det(\cdot),|\cdot|,\|\cdot\|$ & trace, determinant, scalar modulus, Euclidean norm \\
$\vecop(\cdot),\ \otimes$ & vectorization, Kronecker product \\
$\Real\{\cdot\},\ \Imag\{\cdot\}$ & real and imaginary parts \\
$\E\{\cdot\}$ & statistical expectation \\
$\mat{I}_N$ & $N\times N$ identity matrix \\
$\CN(\vect{\mu},\mat{\Sigma})$ & circularly-symmetric complex Gaussian \\
\midrule
\multicolumn{2}{l}{\emph{System dimensions}}\\
$N_t,\ N_r$ & number of transmit, receive antennas \\
$K$ & number of communication users \\
$M$ & number of sensing targets \\
$L$ & number of signal snapshots / symbols \\
\midrule
\multicolumn{2}{l}{\emph{Signals and channels}}\\
$\vect{x}(t),\ \mat{X}$ & transmit signal vector, signal matrix \\
$s_k[\ell],\ \vect{s}[\ell]$ & data symbol of user $k$, symbol vector \\
$\mat{W},\ \vect{w}_k$ & precoding matrix, beamformer for user $k$ \\
$\mat{R}_x$ & transmit signal covariance, $\mat{R}_x=\tfrac{1}{L}\mat{X}\mat{X}^{\herm}$ \\
$\mat{H},\ \vect{h}_k$ & multiuser ($K\times N_t$) channel matrix, user-$k$ channel \\
$\vect{n},\ \sigma^2$ & noise vector, noise power \\
$\mat{\Phi}$ & RIS phase-shift matrix \\
\midrule
\multicolumn{2}{l}{\emph{Sensing parameters}}\\
$\theta,\ \tau,\ \nu$ & target angle, delay (range), Doppler (velocity) \\
$\vect{a}(\theta),\ \vect{b}(\theta)$ & transmit, receive steering vector at angle $\theta$ \\
$\alpha$ & complex target reflection coefficient \\
$\vect{\eta}$ & vector of unknown target parameters \\
$\chi(\tau,\nu)$ & ambiguity function \\
$\FIM(\vect{\eta})$ & Fisher information matrix \\
$\CRB(\cdot)$ & Cram\'er--Rao bound \\
\midrule
\multicolumn{2}{l}{\emph{Waveform parameters}}\\
$N,\ \Delta f,\ B$ & number of subcarriers, subcarrier spacing, bandwidth $B=N\Delta f$ \\
$M_{\mathrm{sym}},\ T_{\mathrm{sym}}$ & OFDM symbols per processing interval, symbol duration \\
$f_c,\ \lambda,\ c$ & carrier frequency, wavelength, speed of light \\
\midrule
\multicolumn{2}{l}{\emph{Performance metrics}}\\
$R,\ C$ & achievable rate, capacity (bits/s/Hz) \\
$\gamma_k$ & SINR of user $k$ \\
$P_d,\ P_{fa}$ & detection, false-alarm probability \\
$D$ & sensing distortion (e.g., estimation MSE) \\
$P_T$ & total transmit power budget \\
\bottomrule
\end{tabular}
\end{table}

\section{The Unified Signal Model}
\label{sec:model}

Everything that follows rests on a common signal model. We adopt the complex-baseband
(equivalent lowpass) representation throughout: a real bandpass signal
$s_{\mathrm{RF}}(t)=\Real\{s(t)e^{j2\pi f_c t}\}$ of carrier frequency $f_c$ is represented by
its complex envelope $s(t)$, which carries all the information and to which all processing is
applied.

\subsection{Transmit Signal}
Consider a dual-function transmitter equipped with a uniform linear array (ULA) of $N_t$
antennas. In a given channel use it emits the baseband vector
$\vect{x}(t)\in\Cset^{N_t}$. Over a block of $L$ discrete samples (or symbols) we collect the
transmitted signal into the matrix
\begin{equation}
\mat{X}=[\vect{x}[1],\vect{x}[2],\dots,\vect{x}[L]]\in\Cset^{N_t\times L}.
\end{equation}
A quantity that will appear repeatedly is the \emph{sample transmit covariance}
\begin{equation}
\mat{R}_x \defeq \frac{1}{L}\mat{X}\mat{X}^{\herm}\in\Cset^{N_t\times N_t},
\label{eq:Rx}
\end{equation}
which is Hermitian positive semidefinite and whose trace equals the average transmit power,
$\tr(\mat{R}_x)=P_T$. As we will see, many of the central metrics of both functions (the
transmit beampattern, the angle-estimation CRB, and covariance-based communication rates) depend
on $\mat{X}$ through $\mat{R}_x$; delay--Doppler sensing additionally depends on the temporal and
spectral structure of the waveform (its ambiguity function, bandwidth, and coherent processing
interval, Section~\ref{sec:radar}), and multiuser rates on the decomposition of $\mat{R}_x$ into
the individual beamformers (Section~\ref{sec:comm}). This shared dependence on a common transmit
resource is the mathematical origin of the ISAC tradeoff.

For a communication-centric system, the transmitted block is generated by linearly precoding
data symbols. Let $\vect{s}[\ell]=[s_1[\ell],\dots,s_K[\ell]]^{\tp}\in\Cset^K$ collect the
symbols intended for the $K$ users at time $\ell$, with
$\E\{\vect{s}[\ell]\vect{s}^{\herm}[\ell]\}=\mat{I}_K$, and let
$\mat{W}=[\vect{w}_1,\dots,\vect{w}_K]\in\Cset^{N_t\times K}$ be the precoding matrix. Then
\begin{equation}
\vect{x}[\ell]=\mat{W}\vect{s}[\ell]=\sum_{k=1}^{K}\vect{w}_k s_k[\ell],
\label{eq:precode}
\end{equation}
and, since $\E\{\vect{s}[\ell]\vect{s}^{\herm}[\ell]\}=\mat{I}_K$, the \emph{statistical}
transmit covariance is
$\E\{\vect{x}[\ell]\vect{x}^{\herm}[\ell]\}=\mat{W}\mat{W}^{\herm}
=\sum_{k=1}^K\vect{w}_k\vect{w}_k^{\herm}$, which the sample covariance~\eqref{eq:Rx}
approaches for sufficiently long blocks $L$. With this understanding we write
$\mat{R}_x=\mat{W}\mat{W}^{\herm}$ throughout; Section~\ref{sec:limits} revisits this
identification, because the finite-$L$ fluctuation of the sample covariance around
$\mat{W}\mat{W}^{\herm}$ is exactly what the deterministic--random tradeoff quantifies.

\subsection{Array Response}
For a ULA with half-wavelength element spacing, a planar wavefront impinging from (or
radiated toward) angle $\theta$ induces the steering vector
\begin{equation}
\vect{a}(\theta)=\big[1,\,e^{j\pi\sin\theta},\,e^{j2\pi\sin\theta},\dots,
e^{j(N_t-1)\pi\sin\theta}\big]^{\tp}\in\Cset^{N_t}.
\label{eq:steer}
\end{equation}
The transmit beampattern, that is, the power radiated toward angle $\theta$, is the quadratic form
\begin{equation}
P(\theta)=\vect{a}^{\herm}(\theta)\,\mat{R}_x\,\vect{a}(\theta),
\label{eq:beampattern}
\end{equation}
which shows explicitly that beampattern design \emph{is} the design of $\mat{R}_x$. This
single identity links the sensing objective (shaping $P(\theta)$ toward targets) to the
communication objective (choosing $\mat{W}$, hence $\mat{R}_x$, for the users).

\subsection{Communication Received Signal}
User $k$ observes the signal through its channel $\vect{h}_k\in\Cset^{N_t}$ (the $k$-th row of
the overall channel $\mat{H}\in\Cset^{K\times N_t}$ being $\vect{h}_k^{\herm}$). Its received sample is
\begin{equation}
\begin{aligned}
y_k[\ell]&=\vect{h}_k^{\herm}\vect{x}[\ell]+n_k[\ell]\\
&=\underbrace{\vect{h}_k^{\herm}\vect{w}_k s_k[\ell]}_{\text{desired}}
+\underbrace{\sum_{i\neq k}\vect{h}_k^{\herm}\vect{w}_i s_i[\ell]}_{\text{interference}}
+\,n_k[\ell],
\end{aligned}
\label{eq:commrx}
\end{equation}
with $n_k[\ell]\sim\CN(0,\sigma^2)$. The decomposition into a desired term and a
multiuser-interference term in~\eqref{eq:commrx} is what makes the received
signal-to-interference-plus-noise ratio (SINR), and hence the achievable rate, a function of \emph{all} the beamformers
jointly, a coupling we resolve in Section~\ref{sec:comm}.

\subsection{Sensing Received Signal}
For sensing, the transmitter illuminates the scene and processes the backscattered echoes. In
the common \emph{monostatic} configuration, the transmit and receive arrays are co-located, so
the receiver knows $\mat{X}$ exactly. Consider $M$ point targets, the $m$-th having angle
$\theta_m$, round-trip delay $\tau_m$ (related to range by $r_m=c\tau_m T_s/2$, with $T_s$ the
sampling interval, because $\tau_m$ is measured in samples below), Doppler shift
$\nu_m$ (proportional to radial velocity), and complex reflection coefficient $\alpha_m$
(absorbing radar cross-section and two-way path loss). The receive array observes
\begin{equation}
\vect{y}_s[\ell]=\sum_{m=1}^{M}\alpha_m\,\vect{b}(\theta_m)\vect{a}^{\herm}(\theta_m)\,
\vect{x}[\ell-\tau_m]\,e^{j2\pi\nu_m \ell}+\vect{z}[\ell],
\label{eq:sensrx}
\end{equation}
where $\vect{a}(\theta)$ and $\vect{b}(\theta)\in\Cset^{N_r}$ are the transmit and receive
steering vectors (the latter defined analogously to~\eqref{eq:steer} over the receive array;
the Hermitian $\vect{a}^{\herm}(\theta)$ matches the beampattern~\eqref{eq:beampattern}, since
the power reaching angle $\theta$ is $\E\{|\vect{a}^{\herm}(\theta)\vect{x}[\ell]|^2\}=P(\theta)$),
the delay $\tau_m$ is expressed in samples and the Doppler shift $\nu_m$ in cycles per sample
(i.e., normalized to the sampling rate), and $\vect{z}[\ell]$ collects noise and clutter. In
compact matrix form over the block, with a
single dominant target for clarity,
\begin{equation}
\mat{Y}_s=\alpha\,\vect{b}(\theta)\vect{a}^{\herm}(\theta)\,\mat{X}(\tau,\nu)+\mat{Z},
\label{eq:sensmatrix}
\end{equation}
where $\mat{X}(\tau,\nu)$ denotes the transmitted block delayed by $\tau$ and Doppler-shifted
by $\nu$. The estimation task is to recover the target parameters
$\vect{\eta}=[\theta,\tau,\nu,\Real\{\alpha\},\Imag\{\alpha\}]^{\tp}$ from $\mat{Y}_s$; the
detection task is to decide whether a target is present at all. These two tasks, developed
next, constitute the sensing half of ISAC.

\subsection{Noise, Clutter, and the Bistatic Case}
Two modeling remarks will matter later. First, the disturbance $\vect{z}[\ell]$ is often
richer than white noise: \emph{clutter} (unwanted reflections from the environment) is
typically colored and can dominate the noise floor, so the whitening of $\vect{z}$ enters the
achievable estimation accuracy~\cite{chiriyath2015effect}. Second, in a \emph{bistatic} or
\emph{multistatic} configuration the transmitter and receiver(s) are separated, which
complicates synchronization (the receiver may not know the transmit phase) but provides
geometric diversity that improves sensing; networked ISAC (Section~\ref{sec:arch}) exploits
exactly this diversity. Unless stated otherwise we develop the monostatic case, which is the
standard setting for the foundational results.

\section{Radar Sensing Foundations}
\label{sec:radar}

This section builds the estimation- and detection-theoretic toolkit that quantifies how well a
waveform can sense. A reader familiar with radar may skim to Section~\ref{sec:comm}.

\subsection{Matched Filtering and the Ambiguity Function}
\begin{figure*}[!t]
\centering
\includegraphics[width=0.92\textwidth]{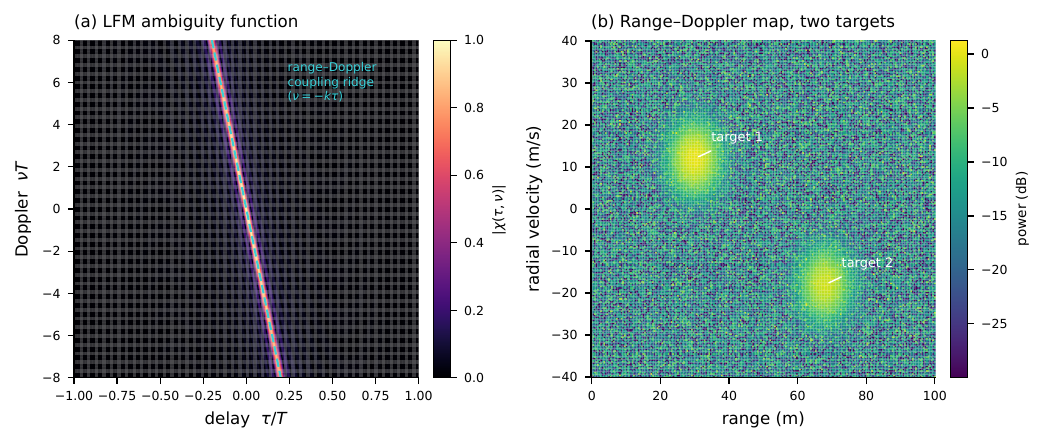}
\caption{(a) Ambiguity function $|\chi(\tau,\nu)|$ of a linear-FM pulse; the diagonal ridge is
range--Doppler coupling. (b) A range--Doppler map formed by two-dimensional matched filtering of
an echo containing two targets, which appear as peaks at their respective range and radial
velocity. Range and velocity resolution are set by the waveform bandwidth and coherent
processing interval, respectively.}
\label{fig:ambiguity}
\end{figure*}

Consider first a single-antenna pulse $x(t)$ of energy $E_x=\int|x(t)|^2\,dt$, reflected by one
target with delay $\tau$ and Doppler $\nu$, so that the noiseless echo is
$r(t)=\alpha\,x(t-\tau)e^{j2\pi\nu t}$. The optimal (SNR-maximizing) receiver for a known signal
in white Gaussian noise is the \emph{matched filter}, which correlates the received signal
against delayed, Doppler-shifted replicas of the transmit waveform. The output as a function of
the hypothesized delay $\tilde\tau$ and Doppler $\tilde\nu$ is governed by the waveform's
\emph{ambiguity function}
\begin{equation}
\chi(\tau,\nu)\defeq\int_{-\infty}^{\infty}x(t)\,x^{*}(t-\tau)\,e^{j2\pi\nu t}\,dt,
\label{eq:ambiguity}
\end{equation}
which measures the response of the matched filter to a target displaced by $(\tau,\nu)$ from the
hypothesis. Its key properties follow directly from~\eqref{eq:ambiguity}:
\begin{enumerate}
\item \emph{Maximum at the origin:} $|\chi(\tau,\nu)|\le|\chi(0,0)|=E_x$, so the response peaks
at the true delay--Doppler.
\item \emph{Constant volume:} $\iint|\chi(\tau,\nu)|^2\,d\tau\,d\nu=E_x^2$, independent of the
waveform. This is the radar uncertainty principle: sharpening the peak in one region necessarily
raises sidelobes elsewhere; resolution cannot be created, only redistributed.
\item \emph{Resolution:} the widths of the central peak in $\tau$ and $\nu$ set the range and
velocity resolution, and are inversely proportional to the waveform's bandwidth and duration,
respectively.
\end{enumerate}
Fig.~\ref{fig:ambiguity}(a) shows the ambiguity function of a linear frequency-modulated (LFM,
or chirp) pulse, whose characteristic sheared ridge illustrates \emph{range--Doppler coupling}:
a delay error masquerades as a Doppler error along the ridge. The ambiguity function is the
central tool for waveform quality, and its ISAC relevance is direct: an ISAC waveform must keep
a sharp ambiguity peak (for sensing) while carrying information (for communication), a tension we
quantify in Section~\ref{sec:limits}~\cite{wang2025optimal}.

\subsection{Maximum-Likelihood Estimation}
Return to the MIMO model~\eqref{eq:sensmatrix}. Stacking the observation as
$\vect{y}=\vecop(\mat{Y}_s)$ and collecting the unknowns into
$\vect{\eta}=[\theta,\tau,\nu,\Real\{\alpha\},\Imag\{\alpha\}]^{\tp}$, the noise
$\vect{z}\sim\CN(\vect{0},\sigma^2\mat{I})$ makes the observation Gaussian with mean
$\vect{\mu}(\vect{\eta})$. The maximum-likelihood (ML) estimator maximizes the log-likelihood
\begin{equation}
\hat{\vect{\eta}}_{\mathrm{ML}}=\arg\max_{\vect{\eta}}\;
-\frac{1}{\sigma^2}\big\|\vect{y}-\vect{\mu}(\vect{\eta})\big\|^2,
\label{eq:ml}
\end{equation}
i.e., it finds the target parameters whose predicted echo best matches the observation. This is exactly
the matched-filter search of the previous subsection, now generalized to the angle dimension via
the steering vectors. The ML estimator is asymptotically efficient, meaning that for long observations or high SNR its
error covariance attains the Cram\'er--Rao bound, which we derive next.

\subsection{The Fisher Information Matrix and the Cram\'er--Rao Bound}
The Cram\'er--Rao bound (CRB) lower-bounds the covariance of \emph{any} unbiased
estimator~\cite{kay1993} and is
the standard metric for ISAC sensing accuracy. For an observation with likelihood
$p(\vect{y};\vect{\eta})$ (Gaussian or not), the Fisher information matrix (FIM) has entries
\begin{equation}
[\FIM(\vect{\eta})]_{ij}
=-\,\E\!\left\{\frac{\partial^2\ln p(\vect{y};\vect{\eta})}{\partial\eta_i\,\partial\eta_j}\right\}.
\label{eq:fim-def}
\end{equation}
For the mean-parameterized complex-Gaussian model $\vect{y}\sim\CN(\vect{\mu}(\vect{\eta}),
\sigma^2\mat{I})$, the Slepian--Bangs formula reduces~\eqref{eq:fim-def} to the convenient form
\begin{equation}
[\FIM(\vect{\eta})]_{ij}
=\frac{2}{\sigma^2}\,\Real\!\left\{
\frac{\partial\vect{\mu}^{\herm}}{\partial\eta_i}\,
\frac{\partial\vect{\mu}}{\partial\eta_j}\right\}.
\label{eq:slepian}
\end{equation}
The CRB is then the inverse of the FIM: for any unbiased estimator $\hat{\vect{\eta}}$,
\begin{equation}
\E\big\{(\hat{\vect{\eta}}-\vect{\eta})(\hat{\vect{\eta}}-\vect{\eta})^{\tp}\big\}
\succeq \FIM^{-1}(\vect{\eta})\defeq\CRB(\vect{\eta}),
\label{eq:crb}
\end{equation}
where $\succeq$ denotes the positive-semidefinite (Loewner) ordering. In particular, the variance
of the $i$-th parameter is bounded by the corresponding diagonal entry,
$\mathrm{var}(\hat\eta_i)\ge[\FIM^{-1}(\vect{\eta})]_{ii}$.

\begin{figure}[!t]
\centering
\includegraphics[width=0.86\columnwidth, trim={0cm 0cm 0cm 0.6cm},clip]{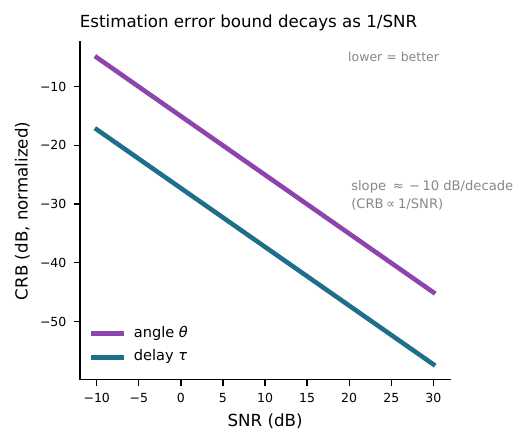}
\caption{The Cram\'er--Rao bound on target-parameter estimation decays inversely with SNR
($\propto1/\mathrm{SNR}$, i.e., $-10$~dB per decade), shown here for angle and delay estimation
in normalized units, so only the slope and the relative offset are meaningful. Because the Fisher information depends on the transmit covariance
$\mat{R}_x$, the same design variable that sets the communication rate also sets these bounds.}
\label{fig:crb}
\end{figure}

Two features of~\eqref{eq:slepian}--\eqref{eq:crb} are worth emphasizing for ISAC. First, the FIM
scales with $1/\sigma^2$, so the CRB scales with $\sigma^2$: the estimation-error bound is
inversely proportional to SNR, falling by $10$~dB for every $10$~dB increase in SNR, as illustrated in
Fig.~\ref{fig:crb}. Second, and crucially, the derivatives
$\partial\vect{\mu}/\partial\eta_j$ depend on the transmitted signal through $\mat{X}$, so the FIM
is a function of the transmit covariance $\mat{R}_x$~\eqref{eq:Rx}. For angle estimation with a
MIMO array, for instance, the CRB takes the schematic form
\begin{equation}
\CRB(\theta)\propto\frac{\sigma^2}{L\,|\alpha|^2\,\dot{\vect{a}}^{\herm}(\theta)\,
\mat{R}_x\,\dot{\vect{a}}(\theta)},
\label{eq:crb-angle}
\end{equation}
where $\dot{\vect{a}}(\theta)=\partial\vect{a}/\partial\theta$. (The exact single-target angle
CRB additionally involves a projection term that accounts for the coupling between $\theta$ and
the unknown nuisance parameters (notably the complex reflection coefficient $\alpha$) and for
the receive array; \eqref{eq:crb-angle} retains the dominant factor to expose the dependence on
$\mat{R}_x$.) Equation~\eqref{eq:crb-angle} is
the pivotal link: \emph{the very same $\mat{R}_x$ that determines the communication rate also
determines the sensing CRB}. Minimizing the CRB and maximizing the rate are therefore two
objectives over one shared variable; this is the essence of ISAC, made precise in
Section~\ref{sec:limits}~\cite{liu2022cram}.

\subsection{Detection Theory}
Estimation presumes a target is present; \emph{detection} decides whether it is. This is a binary
hypothesis test between $\mathcal{H}_0$ (target absent, observation is noise/clutter only) and
$\mathcal{H}_1$ (target present). The Neyman--Pearson lemma states that, to maximize the
probability of detection $P_d$ for a fixed probability of false alarm $P_{fa}$, one should compare
the likelihood ratio to a threshold:
\begin{equation}
\Lambda(\vect{y})=\frac{p(\vect{y}\mid\mathcal{H}_1)}{p(\vect{y}\mid\mathcal{H}_0)}
\underset{\mathcal{H}_0}{\overset{\mathcal{H}_1}{\gtrless}}\;\gamma_{\mathrm{th}},
\label{eq:np}
\end{equation}
with the threshold $\gamma_{\mathrm{th}}$ chosen to meet the $P_{fa}$ constraint. For a known
waveform in white Gaussian noise with an unknown complex amplitude, the generalized
likelihood-ratio form of this test reduces to
comparing the squared magnitude of the matched-filter output against a
threshold~\cite{kay1998}. The resulting tradeoff between
$P_d$ and $P_{fa}$, traced as the threshold varies, is the receiver operating characteristic
(ROC); its shape improves with SNR and with coherent integration over the $L$ samples. In a
constant-false-alarm-rate (CFAR) detector, the threshold is set adaptively from the surrounding
data so that $P_{fa}$ is held fixed as the clutter level varies. Detection and estimation together
constitute the sensing performance that ISAC must deliver alongside communication.

\section{Communication Foundations}
\label{sec:comm}

We now assemble the communication-theoretic results that the ISAC limits require. A reader
versed in wireless communications may skim to Section~\ref{sec:limits}.

\subsection{Shannon Capacity}
For a single-antenna additive white Gaussian noise (AWGN) channel $y=hx+n$ with
$n\sim\CN(0,\sigma^2)$ and transmit-power constraint $\E\{|x|^2\}\le P_T$, the channel capacity
(the maximum rate at which information can be conveyed reliably) is given by the
celebrated Shannon formula~\cite{shannon1948}
\begin{equation}
C=\log_2\!\big(1+\gamma\big)\quad\text{bits/s/Hz},\qquad
\gamma=\frac{|h|^2 P_T}{\sigma^2},
\label{eq:shannon}
\end{equation}
where $\gamma$ is the receive SNR. Note that~\eqref{eq:shannon} is a \emph{spectral efficiency}
(bits/s/Hz): the total rate over a bandwidth $B$ is $B\log_2(1+\gamma)$, which grows only
logarithmically with SNR but linearly with $B$ when the SNR is held fixed (i.e., when the
transmit power scales with the bandwidth; for a fixed total power the SNR falls as $B$ grows and
the rate saturates). This favorable bandwidth scaling is one reason ISAC pursues the large
bandwidths of the millimeter-wave and terahertz bands (Section~\ref{sec:arch}).

\subsection{MIMO Capacity and Water-Filling}
With $N_t$ transmit and $N_r$ receive antennas and channel $\mat{H}\in\Cset^{N_r\times N_t}$
(in this subsection only, $\mat{H}$ denotes a single-user point-to-point channel and $N_r$ the
receive antennas of that user), the
received vector is $\vect{y}=\mat{H}\vect{x}+\vect{n}$. Under a transmit-covariance constraint
$\mat{Q}=\E\{\vect{x}\vect{x}^{\herm}\}\succeq\vect{0}$, $\tr(\mat{Q})\le P_T$, the MIMO capacity
is~\cite{telatar1999}
\begin{equation}
C=\max_{\mat{Q}\succeq\vect{0},\,\tr(\mat{Q})\le P_T}
\;\log_2\det\!\Big(\mat{I}_{N_r}+\tfrac{1}{\sigma^2}\mat{H}\mat{Q}\mat{H}^{\herm}\Big).
\label{eq:mimocap}
\end{equation}
Diagonalizing the channel by its singular value decomposition
$\mat{H}=\mat{U}\mat{\Sigma}\mat{V}^{\herm}$ with singular values $\{\sigma_i\}$ (not to be
confused with the noise power $\sigma^2$) turns the MIMO
channel into a set of parallel scalar subchannels, and the optimal power allocation across them is
\emph{water-filling}:
\begin{equation}
p_i^{\star}=\Big(\mu_{\mathrm{w}}-\frac{\sigma^2}{\sigma_i^2}\Big)^{+},\qquad
\sum_i p_i^{\star}=P_T,
\label{eq:waterfill}
\end{equation}
where $(\cdot)^{+}=\max(\cdot,0)$ and the water level $\mu_{\mathrm{w}}$ is chosen to meet the
power budget (the subscript distinguishes it from the mean vector $\vect{\mu}(\vect{\eta})$
of Section~\ref{sec:radar}).
The resulting $\mat{Q}^{\star}=\mat{V}\diag(p_1^{\star},\dots)\mat{V}^{\herm}$ is the
capacity-achieving transmit covariance. The parallel between the communication covariance $\mat{Q}$
in~\eqref{eq:mimocap} and the sensing covariance $\mat{R}_x$ in~\eqref{eq:crb-angle} is not a
coincidence: in ISAC they are the \emph{same} matrix, which is exactly why the two functions
compete.

\subsection{Multiuser Precoding and Achievable Rate}
In the multiuser downlink of~\eqref{eq:commrx}, treating multiuser interference as noise, the SINR
of user $k$ under precoders $\{\vect{w}_i\}$ is
\begin{equation}
\gamma_k=\frac{|\vect{h}_k^{\herm}\vect{w}_k|^2}
{\displaystyle\sum_{i\neq k}|\vect{h}_k^{\herm}\vect{w}_i|^2+\sigma^2},
\label{eq:sinr}
\end{equation}
and its achievable rate is $R_k=\log_2(1+\gamma_k)$. The sum rate is $\sum_k R_k$. Classical
designs choose $\{\vect{w}_k\}$ to, e.g., maximize the minimum SINR subject to a power budget, or
minimize transmit power subject to per-user SINR targets $\gamma_k\ge\Gamma_k$:
\begin{equation}
\min_{\{\vect{w}_k\}}\;\sum_k\|\vect{w}_k\|^2
\quad\text{s.t.}\quad \gamma_k\ge\Gamma_k,\ \forall k.
\label{eq:power-min}
\end{equation}
Problem~\eqref{eq:power-min}, though not convex as written, admits a well-known convex reformulation
via second-order cone or semidefinite programming~\cite{wiesel2006linear,luo2010sdr}. In ISAC, this same machinery is extended by
appending a sensing constraint or objective in the shared variable $\mat{R}_x=\sum_k\vect{w}_k
\vect{w}_k^{\herm}$, as we develop in Section~\ref{sec:beamforming}. With the sensing toolkit of
Section~\ref{sec:radar} and the communication toolkit of this section in hand, we can now state the
fundamental limits that govern their integration.

\section{Fundamental Limits of ISAC}
\label{sec:limits}

We can now make precise the sense in which sensing and communication trade off. Three
complementary characterizations have emerged, each illuminating a different facet of the same
underlying tension~\cite{liu2023fundamental,xiong2023fundamental}.

\subsection{The CRB--Rate Region}
\begin{figure}[!t]
\centering
\includegraphics[width=0.99\columnwidth, trim={0cm 0cm 0cm 0.6cm},clip]{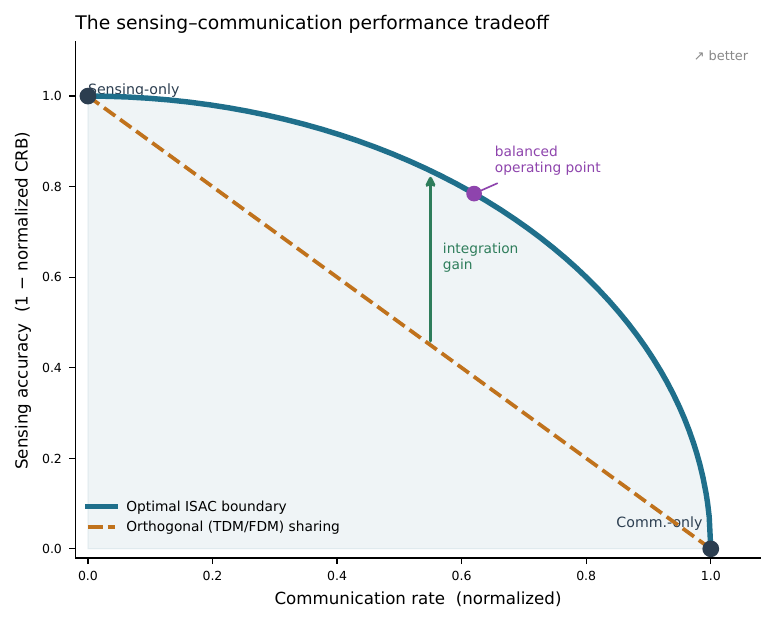}
\caption{The sensing--communication tradeoff as the Pareto boundary of the CRB--rate region.
For readability, sensing accuracy is plotted as $1-$normalized CRB, so that up and to the right
is better; the region in~\eqref{eq:crbregion} is stated in terms of the CRB itself.
The optimally integrated design (solid) is obtained by solving~\eqref{eq:crbrate} and
dominates orthogonal time/frequency sharing between the sensing-only and communication-only
corner points (dashed). The vertical separation is the integration gain. Because time
sharing is linear in the rate and in the Fisher information rather than in the CRB, the dashed
line is schematic in these coordinates.}
\label{fig:tradeoff}
\end{figure}

The most operationally direct characterization pairs the communication rate with the sensing CRB,
both expressed as functions of the shared transmit covariance $\mat{R}_x$ (with the mild abuse of
notation, cf.\ Section~\ref{sec:model}, that the multiuser rate also depends on the decomposition
of $\mat{R}_x$ into the individual beamformers). Recall that the sum rate is (via~\eqref{eq:sinr})
\begin{equation}
R(\mat{R}_x)=\sum_{k}\log_2(1+\gamma_k),
\end{equation}
while the sensing accuracy is measured by a scalarization of the CRB, e.g.\ $\CRB(\theta)$
in~\eqref{eq:crb-angle}. The set of simultaneously achievable pairs
\begin{equation}
\begin{aligned}
\mathcal{C}=\big\{(R,\CRB):\ &\exists\,\mat{R}_x\succeq\vect{0},\ \tr(\mat{R}_x)\le P_T,\\
&R\le R(\mat{R}_x),\ \CRB\ge\CRB(\mat{R}_x)\big\}
\end{aligned}
\label{eq:crbregion}
\end{equation}
defines the \emph{CRB--rate region}, and its Pareto boundary (the pairs at which neither
metric can be improved without degrading the other) is traced by solving,
for each rate target $R_0$,
\begin{equation}
\min_{\mat{R}_x\succeq\vect{0}}\ \CRB(\mat{R}_x)
\quad\text{s.t.}\quad R(\mat{R}_x)\ge R_0,\ \tr(\mat{R}_x)\le P_T.
\label{eq:crbrate}
\end{equation}
Sweeping $R_0$ produces the boundary sketched in Fig.~\ref{fig:tradeoff}. The key qualitative
result, established for canonical Gaussian settings
in~\cite{liu2022cram,xiong2023fundamental} and~\cite{hua2024mimo}, is
that this boundary bulges outward from, and hence dominates, the straight line obtained by time- or
frequency-sharing between a sensing-optimal and a communication-optimal covariance: an integrated
design that reuses one signal for both functions outperforms orthogonally splitting the resource.
The precise shape of the boundary and the size of this gap depend on the channel model, the
sensing metric and its scalarization, and the feasible waveform and beamforming set. The vertical
gap between the two curves in Fig.~\ref{fig:tradeoff} is the \emph{integration gain}, the
quantitative payoff of ISAC.

\subsection{The Capacity--Distortion Function}
A more information-theoretic viewpoint models ISAC as joint communication and \emph{state
estimation} over a state-dependent channel. The transmitter sends a message while the channel
carries an unknown state $\vect{\eta}$ (the target parameters), which the transmitter (or a
co-located sensing receiver) estimates from the backscatter with distortion
$D=\E\{d(\vect{\eta},\hat{\vect{\eta}})\}$ for a distortion measure $d(\cdot,\cdot)$. The
fundamental limit is the \emph{capacity--distortion function}
\begin{equation}
C(D)=\max_{p(x):\,\E\{d\}\le D}\ I(X;Y),
\label{eq:capdist}
\end{equation}
the largest communication rate (mutual information $I(X;Y)$) achievable while holding the sensing
distortion below $D$~\cite{chen2023general,paul2016joint}; this formulation was introduced
for memoryless channels in~\cite{kobayashi2018joint} and generalized
in~\cite{ahmadipour2024information}. Expression~\eqref{eq:capdist} is a
tutorial-level schematic: the precise formulation depends on the state-dependent channel model,
the side information available to the transmitter and the estimator, and whether the distortion
constraint is imposed on average or per realization; general capacity--distortion formulas
covering these variations are developed in~\cite{chen2023general}. The function $C(D)$ is non-decreasing
and concave in $D$: allowing more sensing distortion frees the input distribution to carry more
information. Its two extremes recover familiar quantities. At $D\to\infty$ (sensing abandoned),
$C(D)$ reaches the ordinary channel capacity; at the minimum feasible distortion $D_{\min}$
(communication subordinated to sensing), the input is restricted to the
estimation-optimal distributions, which in many models are deterministic or constant-modulus
signals.

\subsection{The Deterministic--Random Tradeoff}
The capacity--distortion picture exposes a structural tension that the third characterization makes
explicit~\cite{liu2023fundamental}. Sensing performance generally favors \emph{deterministic}
signals: a known, carefully shaped waveform yields a predictable ambiguity function and stable
estimation performance. Communication performance, by contrast, requires \emph{random} signals: it is precisely
the entropy, or unpredictability, of the transmitted signal that conveys information, and the
capacity-achieving input of the Gaussian channel is Gaussian, i.e., maximally random for a
given power. An ISAC waveform
must therefore be simultaneously predictable enough to sense well and random enough to communicate
well. This \emph{deterministic--random tradeoff} can be made precise through the sample
covariance of Section~\ref{sec:model}. The CRB~\eqref{eq:crb-angle} depends on the \emph{sample}
covariance $\mat{R}_x=\frac{1}{L}\mat{X}\mat{X}^{\herm}$; random data make $\mat{R}_x$
fluctuate around $\mat{W}\mat{W}^{\herm}$, and because the CRB is convex in $\mat{R}_x$, this
fluctuation raises the \emph{expected} CRB (by Jensen's inequality). Sensing is therefore best
served by signals whose sample covariance is deterministic, for instance scaled semi-unitary
probing blocks, whereas $I(X;Y)$ is maximized by Gaussian signaling, whose sample covariance
fluctuates; the penalty shrinks as the block length $L$ grows~\cite{xiong2023fundamental}. The optimal ISAC input
distribution interpolates between these poles, and where a given system sits along the interpolation
is a design choice, not a fixed point~\cite{xiong2023fundamental,xiong2024generalized}.

\subsection{Networked and Multi-Target Extensions}
The single-link bounds above extend to networks of cooperating nodes. When multiple
transmit--receive pairs jointly sense the $M$ targets, the FIM aggregates the information contributed by each
transmitter--receiver geometry, and the CRB reflects the resulting spatial diversity, yielding
cooperation gains analogous to distributed MIMO radar~\cite{li2023performance,chen2025distributed}.
A complete characterization of the achievable region for networked, multi-target,
hardware-constrained ISAC remains open (Section~\ref{sec:challenges}), but the single-link results of this
section supply the template: identify the shared resource, express each function's metric as a
function of it, and optimize over the tradeoff. The remaining sections turn from \emph{what is
achievable} to \emph{how to achieve it}.

\section{Waveform Design}
\label{sec:waveform}

The waveform is the primary design degree of freedom in ISAC: it must possess good communication
properties (high rate, robustness) and good sensing properties (a sharp ambiguity function, low
sidelobes). We first derive OFDM radar processing in detail, then summarize the main waveform
families in Table~\ref{tab:waveforms}.

\subsection{OFDM Radar Processing}
\begin{figure*}[!t]
\centering
\includegraphics[width=0.96\textwidth,trim={0.5cm 0.5cm 0.4cm 0.8cm},clip]{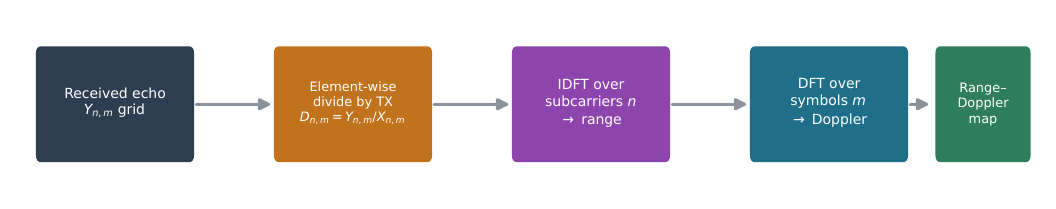}
\caption{OFDM radar processing. Because the transmitter knows the payload symbols, dividing the
received time--frequency grid by the transmitted symbols removes the data modulation and leaves a
two-dimensional sinusoid whose frequencies encode delay and Doppler; a two-dimensional FFT then
produces the range--Doppler map of Fig.~\ref{fig:ambiguity}(b).}
\label{fig:ofdm}
\end{figure*}

Orthogonal frequency-division multiplexing (OFDM) is the dominant communication waveform and a
leading ISAC candidate, because its known time--frequency structure permits an elegant
range--Doppler estimator~\cite{sturm2009novel,zeng2026ofdm}. Let $X_{n,m}$ be the transmitted
complex data symbol on subcarrier $n\in\{0,\dots,N-1\}$ of OFDM symbol
$m\in\{0,\dots,M_{\mathrm{sym}}-1\}$ (we write $M_{\mathrm{sym}}$ to avoid a clash with the
number of targets $M$), with
subcarrier spacing $\Delta f$ and symbol duration $T_{\mathrm{sym}}$ (including the cyclic
prefix). Assuming the round-trip
delay falls within the cyclic prefix (which bounds the directly processable range; see
Section~\ref{sec:example}) and the Doppler-induced inter-carrier interference is
negligible, a target at delay $\tau$ (in seconds, unlike the normalized units
of~\eqref{eq:sensrx}) and Doppler $\nu$ (in hertz) produces the received grid
\begin{equation}
Y_{n,m}=\alpha\,X_{n,m}\,
\underbrace{e^{-j2\pi n\Delta f\tau}}_{\text{delay}\to\text{phase over }n}\,
\underbrace{e^{j2\pi m T_{\mathrm{sym}}\nu}}_{\text{Doppler}\to\text{phase over }m}
+Z_{n,m}.
\label{eq:ofdmrx}
\end{equation}
The essential step is \emph{reciprocal filtering}: because the receiver knows the transmitted
symbols exactly (monostatic sensing), it divides them out element-wise,
\begin{equation}
D_{n,m}\defeq\frac{Y_{n,m}}{X_{n,m}}
=\alpha\,e^{-j2\pi n\Delta f\tau}\,e^{j2\pi m T_{\mathrm{sym}}\nu}+Z'_{n,m},
\label{eq:ofdm-divide}
\end{equation}
which removes the data modulation entirely and leaves a pure two-dimensional complex sinusoid whose
frequencies encode delay (along $n$) and Doppler (along $m$). The target parameters are then read
off by a two-dimensional discrete Fourier transform, namely an inverse DFT along subcarriers to resolve
range and a DFT along symbols to resolve Doppler:
\begin{equation}
\Xi(p,q)=\sum_{m=0}^{M_{\mathrm{sym}}-1}\sum_{n=0}^{N-1}D_{n,m}\,e^{j2\pi np/N}\,e^{-j2\pi mq/M_{\mathrm{sym}}},
\label{eq:ofdm-2dfft}
\end{equation}
whose magnitude peaks at the bin $(p,q)$ corresponding to the target's range and velocity, giving
exactly the range--Doppler map of Fig.~\ref{fig:ambiguity}(b). This processing chain is summarized
in Fig.~\ref{fig:ofdm}. The unambiguous range is set by the subcarrier spacing
($r_{\max}=c/(2\Delta f)$) and the range resolution by the total bandwidth
($\Delta r=c/(2N\Delta f)$); analogously, the velocity resolution is set by the coherent
processing interval ($\Delta v=\lambda/(2M_{\mathrm{sym}}T_{\mathrm{sym}})$ at wavelength $\lambda$) and the
unambiguous velocity by the symbol duration ($|v|<\lambda/(4T_{\mathrm{sym}})$, up to the Doppler
sampling convention). A key drawback is that the division in~\eqref{eq:ofdm-divide} requires
nonzero symbols (or pilot-based regularization) and can enhance noise where $|X_{n,m}|$ is small,
and OFDM's high peak-to-average-power ratio (PAPR) stresses the amplifier.

\subsection{OTFS and Delay--Doppler Waveforms}
In high-mobility scenarios the doubly-dispersive channel spreads OFDM subcarriers, degrading both
functions. Orthogonal time--frequency space (OTFS) modulation instead places information symbols in
the \emph{delay--Doppler} domain, where the channel is sparse and quasi-static, and maps them to the
time--frequency domain for transmission via the inverse symplectic finite Fourier
transform~\cite{hadani2017orthogonal,huang2026novel}. Because the sensing parameters
$(\tau,\nu)$ are themselves delay--Doppler coordinates, OTFS aligns the communication signal domain
with the sensing observation domain, making it a natural ISAC waveform for vehicular and aerial
links, at the cost of higher receiver complexity.

\subsection{Jointly Optimized Waveforms}
The most flexible approach optimizes the transmitted block $\mat{X}$ (equivalently $\mat{R}_x$)
directly against a weighted ISAC objective subject to practical constraints. A representative
formulation trades a communication metric against a sensing metric,
\begin{equation}
\begin{aligned}
\min_{\mat{X}}\ \ &\rho\,\mathcal{L}_{\mathrm{sense}}(\mat{X})
+(1-\rho)\,\mathcal{L}_{\mathrm{comm}}(\mat{X})\\[2pt]
\text{s.t.}\ \ &\tfrac{1}{L}\|\mat{X}\|_F^2\le P_T\quad\text{(power)},\\
&|x_i[\ell]|=\sqrt{P_T/N_t}\quad\text{(constant modulus)},
\end{aligned}
\label{eq:jointwave}
\end{equation}
where $\rho\in[0,1]$ traces the CRB--rate boundary of Section~\ref{sec:limits} (a weighted
sum recovers only the points of the boundary that lie on its convex hull; constraint-based
formulations such as~\eqref{eq:crbrate} recover all of them),
$\mathcal{L}_{\mathrm{sense}}$ might be a beampattern-matching error or the CRB, and
$\mathcal{L}_{\mathrm{comm}}$ a multiuser-interference or rate-gap penalty~\cite{liu2018dual}. The
constant-modulus constraint, which at this level also meets the power budget with equality,
lets the power amplifier operate efficiently near saturation ($0$~dB PAPR) but is non-convex,
so~\eqref{eq:jointwave} is typically solved by manifold optimization, successive convex
approximation, or alternating minimization. Extensions incorporate one-bit converters and
continuous-aperture arrays directly into the constraint set~\cite{lin2026crb,ye2025optimal},
spatial-division designs superimpose the sensing signal in the null space of the
communication channel~\cite{lee2024spatial}, and the achievable PAPR--ambiguity tradeoff bounds
what any such waveform can attain~\cite{wang2025optimal}. Comprehensive reviews survey the full design
space~\cite{hanif2023exploring}.

\begin{table*}[!t]
\centering
\caption{Representative ISAC Waveform Families and Their Characteristics}
\label{tab:waveforms}
\renewcommand{\arraystretch}{1.35}
\begin{tabular}{p{2.6cm} p{3.3cm} p{4.3cm} p{4.3cm}}
\toprule
\textbf{Waveform family} & \textbf{Design origin} & \textbf{Strengths} & \textbf{Limitations} \\
\midrule
OFDM-based & Communication-centric~\cite{sturm2009novel,zeng2026ofdm} &
Mature ecosystem; reciprocal filtering enables efficient 2D-FFT range--Doppler estimation &
High PAPR; noise enhancement in division; range limited by subcarrier spacing and cyclic
prefix \\
OTFS / delay--Doppler & Communication-centric~\cite{hadani2017orthogonal,huang2026novel} &
Robust in high-mobility channels; signal domain matches sensing domain &
Higher receiver complexity; less mature standardization \\
Spread-spectrum / codebook & Radar-centric~\cite{hassanien2016signaling} &
Preserves radar waveform quality; bits via sidelobe or waveform-index modulation &
Low communication rate; suited to auxiliary links only \\
DFT-s-OFDM with chirping & Communication-centric~\cite{liu2026dft} &
Low PAPR for uplink and power-limited nodes; chirp aids sensing &
Sensing resolution constrained by chirp parameters \\
Optimized joint waveform & Joint design~\cite{liu2018dual,wang2025optimal,lee2024spatial} &
Directly traces the CRB--rate frontier; incorporates hardware constraints &
Computationally intensive; needs channel/target priors \\
\bottomrule
\end{tabular}
\end{table*}

\section{Beamforming and Receiver Processing}
\label{sec:beamforming}

Beamforming exploits the spatial degrees of freedom of a multi-antenna array to serve communication
users and illuminate sensing directions simultaneously. This section casts ISAC beamforming as an
explicit optimization problem and outlines the receiver algorithms that close the sensing loop.

\subsection{Transmit Beampattern Design}
Recall from~\eqref{eq:beampattern} that the power radiated toward angle $\theta$ is
$P(\theta)=\vect{a}^{\herm}(\theta)\mat{R}_x\vect{a}(\theta)$. A purely sensing-oriented design
chooses $\mat{R}_x$ so that $P(\theta)$ matches a desired beampattern $P_{\mathrm{des}}(\theta)$ that
concentrates power on the target angles $\{\theta_m\}$, by solving the least-squares fit
\begin{equation}
\begin{aligned}
\min_{\mat{R}_x\succeq\vect{0}}\ &\sum_{g}\big|P_{\mathrm{des}}(\theta_g)-\vect{a}^{\herm}(\theta_g)\mat{R}_x
\vect{a}(\theta_g)\big|^2\\
\text{s.t.}\ &\tr(\mat{R}_x)=P_T,\ [\mat{R}_x]_{ii}=\tfrac{P_T}{N_t},
\end{aligned}
\label{eq:beampattern-match}
\end{equation}
over a grid $\{\theta_g\}$, where the per-antenna power constraint reflects hardware limits. This is
a convex semidefinite program in $\mat{R}_x$, following the classical MIMO-radar probing-signal
design of Stoica \emph{et al.}~\cite{stoica2007probing}, who also optimize a scale factor
multiplying $P_{\mathrm{des}}$; see~\cite{li2007mimo} for a broader treatment of colocated MIMO
radar.

\subsection{Joint ISAC Beamforming}
The ISAC problem couples~\eqref{eq:beampattern-match} with the communication metrics of
Section~\ref{sec:comm} through the shared identity
$\mat{R}_x=\sum_k\vect{w}_k\vect{w}_k^{\herm}$. Two dual formulations are standard. In the first,
communication quality of service is guaranteed and sensing is optimized:
\begin{equation}
\min_{\{\vect{w}_k\}}\ \CRB(\theta)
\quad\text{s.t.}\quad \gamma_k\ge\Gamma_k\ \forall k,\ \ \sum_k\|\vect{w}_k\|^2\le P_T,
\label{eq:isac-crbmin}
\end{equation}
which minimizes the sensing bound subject to per-user SINR floors $\Gamma_k$; this dual-objective beamforming lineage traces back to joint transmit designs for coexisting MU-MIMO communication and MIMO radar~\cite{liu2018mimo,liu2020jointtsp,liu2022cram}. In
the second, sensing quality is guaranteed and communication is optimized:
\begin{equation}
\begin{aligned}
\max_{\{\vect{w}_k\}}\ &\sum_k\log_2(1+\gamma_k)\\
\text{s.t.}\ &\vect{a}^{\herm}(\theta_m)\mat{R}_x\vect{a}(\theta_m)\ge\Gamma_s\ \forall m,\
\sum_k\|\vect{w}_k\|^2\le P_T,
\end{aligned}
\label{eq:isac-ratemax}
\end{equation}
which maximizes sum rate subject to a minimum illumination $\Gamma_s$ toward each target. Sweeping
$\Gamma_k$ in~\eqref{eq:isac-crbmin} traces the CRB--rate boundary of
Fig.~\ref{fig:tradeoff}; sweeping $\Gamma_s$ in~\eqref{eq:isac-ratemax} traces the analogous
rate--illumination boundary, which tracks the CRB--rate boundary only insofar as illumination
power governs the CRB.

Two refinements are worth noting. First, the transmitter may add dedicated sensing streams,
$\vect{x}[\ell]=\mat{W}\vect{s}[\ell]+\vect{x}_0[\ell]$ with
$\E\{\vect{x}_0[\ell]\vect{x}_0^{\herm}[\ell]\}=\mat{R}_0\succeq\vect{0}$, so that
$\mat{R}_x=\sum_k\vect{w}_k\vect{w}_k^{\herm}+\mat{R}_0$. This removes the restriction
$\rank(\mat{R}_x)\le K$, which matters when $K<N_t$ (for example, for extended targets or several
sensing directions), at the cost of extra interference at users that cannot cancel the known
sensing signal~\cite{liu2020jointtsp}. Second, the CRB objective in~\eqref{eq:isac-crbmin} can be
written as a linear matrix inequality in $\mat{R}_x$ through the Schur complement, so it adds no
difficulty beyond that of the SINR constraints~\cite{liu2022cram}.

\subsection{Semidefinite Relaxation}
Problems~\eqref{eq:isac-crbmin}--\eqref{eq:isac-ratemax} are non-convex because the SINR
$\gamma_k$~\eqref{eq:sinr} is a ratio of quadratics in $\vect{w}_k$~\cite{luo2010sdr}. The standard remedy is
\emph{semidefinite relaxation} (SDR): define the rank-one lifted variables
$\mat{W}_k=\vect{w}_k\vect{w}_k^{\herm}\succeq\vect{0}$. The SINR constraint becomes linear in the
$\{\mat{W}_k\}$ once multiplied through by its positive denominator,
\begin{equation}
\frac{\tr(\mat{H}_k\mat{W}_k)}{\sum_{i\neq k}\tr(\mat{H}_k\mat{W}_i)+\sigma^2}\ge\Gamma_k,
\qquad \mat{H}_k\defeq\vect{h}_k\vect{h}_k^{\herm},
\label{eq:sdr}
\end{equation}
and, dropping the non-convex constraint $\rank(\mat{W}_k)=1$, the problem~\eqref{eq:isac-crbmin}
becomes a convex
semidefinite program solvable in polynomial time. (For the sum-rate objective
of~\eqref{eq:isac-ratemax} the relaxed problem remains non-convex and is usually handled by
fractional programming or successive convex approximation.) If the resulting $\{\mat{W}_k^{\star}\}$ happen to
be rank one, the relaxation is tight and $\vect{w}_k^{\star}$ is recovered by eigendecomposition;
otherwise a rank-one approximation (e.g., Gaussian randomization) is used. For the classical
power-minimization problem~\eqref{eq:power-min} the relaxation is known to be
tight~\cite{luo2010sdr}, and for several ISAC formulations a rank-one optimum can be constructed
from the relaxed solution~\cite{liu2020jointtsp}. SDR, together with
successive convex approximation and fractional programming, is the computational backbone of ISAC
beamforming.

\subsection{Hybrid Beamforming}
At millimeter-wave and terahertz frequencies (Section~\ref{sec:arch}), assigning a dedicated
radio-frequency (RF) chain to each antenna is prohibitively costly. \emph{Hybrid beamforming}
factors the precoder $\mat{W}=\mat{F}_{\mathrm{RF}}\mat{F}_{\mathrm{BB}}$ into a high-dimensional
analog part $\mat{F}_{\mathrm{RF}}$ (implemented with phase shifters, hence constant-modulus
entries) and a low-dimensional digital part $\mat{F}_{\mathrm{BB}}$ with far fewer RF chains than
antennas. ISAC-oriented hybrid designs balance the sensing beampattern against the communication
multiplexing gain under this structural constraint~\cite{bayraktar2023hybrid,zhang2025distortion},
and robust or Bayesian variants hedge against imperfect channel and target
knowledge~\cite{zhao2025bayesian,maleki2025dual}.

\subsection{Receiver-Side Estimation and Detection}
At the sensing receiver, the algorithms of Section~\ref{sec:radar} are instantiated. Matched
filtering~\eqref{eq:ambiguity} or the OFDM 2D-FFT~\eqref{eq:ofdm-2dfft} produces a range--Doppler map;
angle is estimated by beamforming across the receive array or by subspace methods such as MUSIC,
which exploit the orthogonality between the signal and noise subspaces of the sample covariance to
achieve super-resolution; and CFAR detection, an adaptive-threshold implementation of the test
in~\eqref{eq:np}, declares targets while controlling the
false-alarm rate~\cite{zhang2021overview}. In the perceptive-network setting, these estimators must
operate within a communication receiver, reusing pilots and reference signals and fusing
measurements across antennas, subcarriers, and time~\cite{zhang2021perceptive,zhang2022enabling}.
Increasingly, learned estimators complement the model-based ones where accurate statistical models
are unavailable~\cite{demirhan2023integrated,mateos2022end}.

\section{Enabling Architectures}
\label{sec:arch}

\begin{figure}[!t]
\centering
\includegraphics[width=0.99\columnwidth, trim={1cm 1.7cm 1cm 0.85cm},clip]{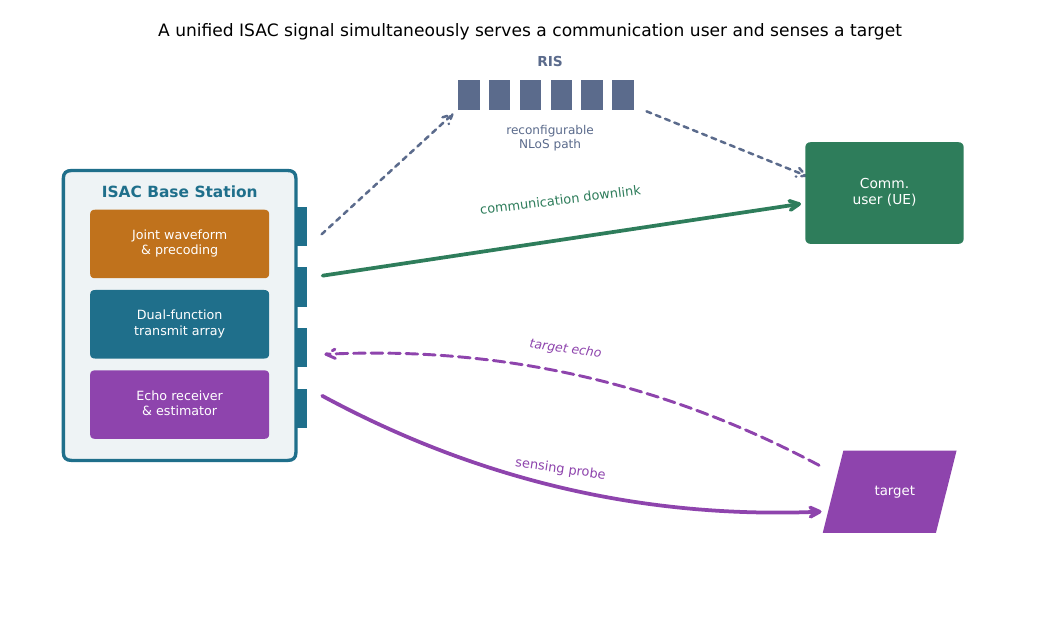}
\caption{A canonical ISAC deployment. A dual-function base station transmits a shared waveform that
both serves communication users and illuminates sensing targets; the backscattered echoes are
processed for target detection and estimation, while a reconfigurable intelligent surface (RIS)
synthesizes an additional controllable propagation path (Section~\ref{sec:arch}).}
\label{fig:architecture}
\end{figure}

The foundations of Sections~\ref{sec:model}--\ref{sec:beamforming} are realized in hardware and
network architectures that scale ISAC from a single link to a perceptive network. A canonical
deployment is sketched in Fig.~\ref{fig:architecture}: a dual-function base station serves
communication users while simultaneously probing and receiving echoes from sensing targets, with a
reconfigurable intelligent surface providing a controllable auxiliary path.

\subsection{Millimeter-Wave and Terahertz ISAC}
Operating at millimeter-wave (mmWave, nominally $30$--$300$~GHz, of which $24.25$--$71$~GHz
forms 3GPP frequency range~2) and terahertz (THz, $0.1$--$10$~THz; the $0.1$--$0.3$~THz
sub-THz range overlaps the upper mmWave band) frequencies
serves both functions at once: the large bandwidths raise communication rate (linearly at a fixed
SNR, via~\eqref{eq:shannon}) and sharpen range resolution ($\Delta r=c/(2B)$), while the large antenna
arrays that combat path loss also narrow the sensing beam and tighten the angle CRB
in~\eqref{eq:crb-angle}~\cite{wei2023signals,wild2023from}. The costs are severe path loss,
blockage sensitivity, and the RF-chain expense that motivates the hybrid beamforming of
Section~\ref{sec:beamforming}. Extra-large arrays introduce near-field effects that further couple
the two functions~\cite{li2023user}.

\subsection{Reconfigurable Intelligent Surfaces}
A reconfigurable intelligent surface (RIS) is a planar array of passive elements that imposes a
tunable phase shift on the impinging wave, synthesizing a controllable propagation path. Model an
RIS with $N_{\mathrm{RIS}}$ elements by its diagonal phase matrix
$\mat{\Phi}=\diag(e^{j\phi_1},\dots,e^{j\phi_{N_{\mathrm{RIS}}}})$; the effective
transmitter--RIS--user channel becomes the \emph{cascade}
\begin{equation}
\vect{h}_{\mathrm{eff}}^{\herm}=\vect{h}_{r}^{\herm}\mat{\Phi}\mat{G}+\vect{h}_{d}^{\herm},
\label{eq:ris}
\end{equation}
where $\mat{G}$ is the transmitter--RIS channel, $\vect{h}_r$ the RIS--user channel, and
$\vect{h}_d$ the direct path~\cite{wu2018intelligent,wu2020smart}. Because $\mat{\Phi}$ multiplies
the sensing and communication channels alike, the RIS becomes a shared design variable much as
$\mat{R}_x$ is at the transmitter, and joint optimization of $(\mat{W},\mat{\Phi})$ can create
sensing paths where none exist, for example around corners and past
blockages~\cite{rivetti2025millimeter,yasmeen2026around,yasmeen2026radar}, aid
positioning~\cite{ta2026reconfigurable,li2024localization,ghazalian2024joint}, and, in
beyond-diagonal and holographic variants, enlarge the tunable
space~\cite{zhang2025beyond,zheng2025beyond,zhang2026holographic,jin2024reconfigurable}.

\subsection{Cell-Free and Networked ISAC}
Rather than a single base station, \emph{cell-free} and \emph{networked} ISAC coordinate many
distributed access points to serve users and sense targets cooperatively. As noted in
Section~\ref{sec:limits}, the multistatic geometry contributes independent Fisher information from
each transmitter--receiver pair, so the aggregate CRB improves with the number and diversity of
links, and macro-diversity mitigates blockage~\cite{andrews2026network,galappaththige2026cell}. The
architecture also raises new resource-allocation and data-handling problems: which nodes sense,
which communicate, and how measurements are fused and transported across the
backhaul~\cite{jadoon2026architectural}. These questions connect ISAC to the broader system and
economic design of the network~\cite{guijarro2024economics}.

\section{Applications, Standardization, and Security}
\label{sec:apps}

\subsection{Applications}
The perceptive network enables a range of services. In \emph{vehicular} networks (V2X), an ISAC
base station or vehicle simultaneously communicates and senses other road users, supporting
collision avoidance and cooperative driving; full-duplex operation lets a node sense while it
transmits~\cite{moulin2024full}. In \emph{low-altitude} and unmanned-aerial-vehicle scenarios, ISAC
provides simultaneous connectivity and airspace surveillance, an emerging pillar of the
low-altitude economy and its security~\cite{ren2026low}. \emph{Localization and environmental
imaging} use the estimated delay, angle, and Doppler of Section~\ref{sec:radar} to position devices
and reconstruct the surroundings, often RIS-aided~\cite{li2024localization,ghazalian2024joint}.
Rate-splitting multiple access has been applied to manage the interference among users and the
sensing function in ISAC for emergency unmanned-aerial-vehicle systems~\cite{yao2024coordinated}.

\subsection{Standardization}
ISAC has moved from research into standards. Recommendation ITU-R M.2160 lists integrated sensing and communication among the six
usage scenarios of IMT-2030~\cite{itu2023m2160}. Within the 3rd Generation Partnership Project
(3GPP), the Release-19 feasibility study TR~22.837 defined sensing use cases and their service
requirements~\cite{tr22837}, and a parallel radio-access study extended the TR~38.901 channel
model to sensing: the channel is modeled as the sum of a target component and a background
component, and six sensing modes are covered, namely base-station (TRP) monostatic, UE
monostatic, and TRP--TRP, TRP--UE, UE--TRP, and UE--UE bistatic~\cite{tr38901}. Industry
roadmaps place ISAC as a native 6G capability~\cite{wei2023signals,wild2023from}. Standardization must reconcile the
sensing requirements derived here (waveform structure, reference-signal density, duplexing) with
the installed communication base, and defines the interfaces by which sensing results are exposed
as a network service.

\subsection{Security and Privacy}
Sensing turns the network into a sensor, which raises security and privacy concerns absent from
communication alone. Two threat classes dominate. First, the sensing signal illuminates the
environment and may be \emph{eavesdropped} or expose private information about sensed subjects;
physical-layer security techniques shape the transmit covariance $\mat{R}_x$ to degrade an
eavesdropper's channel while preserving the legitimate rate and the sensing
function~\cite{welling2024low,welling2024transmitter,su2025integrating}. Second, sensing itself can
be \emph{attacked}, for example by spoofed echoes or by a mobile adversary, which motivates
robust estimation and
authentication of the sensed returns~\cite{mamaghani2025securing}. A recurring theme is that
security must be designed into ISAC from the waveform and beamformer outward, not bolted on
afterward~\cite{su2025integrating}.

\section{A Worked MIMO-OFDM Design Example}
\label{sec:example}

To consolidate the machinery, we trace a single end-to-end ISAC design from signal model to
tradeoff. Consider a base station with $N_t=16$ transmit antennas and an OFDM waveform of $N=1024$
subcarriers spaced $\Delta f=120$~kHz, serving $K=4$ single-antenna users while sensing one target
at angle $\theta_0$. We take a carrier frequency $f_c=28$~GHz ($\lambda\approx1.07$~cm) and a
coherent processing interval of $M_{\mathrm{sym}}=112$ OFDM symbols (eight 14-symbol slots, about
$1$~ms), each of duration $T_{\mathrm{sym}}\approx8.92~\mu$s including a normal cyclic prefix of
$T_{\mathrm{CP}}\approx0.59~\mu$s. The co-located receive array is assumed to match the transmit
array ($N_r=16$).

\emph{Step 1: Signal model.} The transmitted block is $\vect{x}[\ell]=\mat{W}\vect{s}[\ell]$
as in~\eqref{eq:precode}, so the design variables are the beamformers
$\{\vect{w}_k\}_{k=1}^{4}$, and everything downstream depends on them through
$\mat{R}_x=\sum_k\vect{w}_k\vect{w}_k^{\herm}$.

\emph{Step 2: Communication metric.} Each user's SINR is $\gamma_k$ from~\eqref{eq:sinr} and its
rate $R_k=\log_2(1+\gamma_k)$; the system sum rate is $R=\sum_{k=1}^4 R_k$.

\emph{Step 3: Sensing metric.} Illuminating $\theta_0$, the angle estimation bound is
$\CRB(\theta_0)\propto\sigma^2/\!\big(L|\alpha|^2\dot{\vect{a}}^{\herm}(\theta_0)\mat{R}_x
\dot{\vect{a}}(\theta_0)\big)$ from~\eqref{eq:crb-angle}; sharper illumination toward $\theta_0$
lowers the bound.

\emph{Step 4: Waveform processing.} Along each spatial stream, the OFDM echo is processed by the
reciprocal-filter-plus-2D-FFT chain~\eqref{eq:ofdm-divide}--\eqref{eq:ofdm-2dfft}, yielding the
range--Doppler map from which delay and Doppler are read; the $16$-antenna array resolves
$\theta_0$. Substituting the numerology into the expressions of Section~\ref{sec:waveform}
gives a bandwidth $B=N\Delta f=122.88$~MHz and a range resolution $\Delta r=c/(2B)\approx1.22$~m;
an unambiguous range $c/(2\Delta f)\approx1.25$~km, although the cyclic-prefix assumption
behind~\eqref{eq:ofdmrx} limits directly processable echoes to $cT_{\mathrm{CP}}/2\approx88$~m
unless longer delays are handled explicitly; a velocity resolution
$\lambda/(2M_{\mathrm{sym}}T_{\mathrm{sym}})\approx5.4$~m/s; and an unambiguous velocity
$|v|<\lambda/(4T_{\mathrm{sym}})\approx300$~m/s when every symbol is used for sensing. If only
periodically inserted reference signals are used, their repetition period replaces
$T_{\mathrm{sym}}$ and the unambiguous velocity shrinks accordingly. With half-wavelength spacing,
the $16$-element array separates directions roughly $2/N_t$ apart in $\sin\theta$, about
$7^\circ$ at broadside.

\emph{Step 5: Joint optimization.} We solve the CRB-minimizing
program~\eqref{eq:isac-crbmin}: minimize $\CRB(\theta_0)$ subject to $\gamma_k\ge\Gamma$ for all
four users and $\sum_k\|\vect{w}_k\|^2\le P_T$. Applying the SDR of~\eqref{eq:sdr}, we lift to
$\{\mat{W}_k\succeq\vect{0}\}$, obtain a convex SDP, solve it, and recover the beamformers.

\emph{Step 6: Tradeoff.} Sweeping the SINR floor $\Gamma$ traces a boundary of the kind
sketched in
Fig.~\ref{fig:tradeoff}: as $\Gamma$ rises, more power is committed to the users, $\CRB(\theta_0)$
increases, and the achieved $(R,\CRB)$ pair walks along the Pareto boundary. The gap between this
integrated curve and orthogonal sharing is the integration gain the design delivers. This single
example exercises every layer of the tutorial (model, sensing bound, communication rate, waveform
processing, optimization, and tradeoff) and serves as a template for most ISAC system designs.

\section{Open Challenges and Conclusion}
\label{sec:challenges}

\subsection{Open Challenges}
Several problems remain open for the reader entering the field.
\begin{itemize}
\item \emph{Fundamental limits at scale.} The single-link CRB--rate and capacity--distortion results
of Section~\ref{sec:limits} are not yet matched by a complete theory for networked, multi-target,
hardware-constrained ISAC~\cite{andrews2026network,liu2023fundamental}.
\item \emph{Full-duplex and self-interference.} Monostatic sensing while transmitting requires
canceling self-interference far below the noise floor~\cite{moulin2024full}.
\item \emph{Synchronization for bistatic and networked sensing.} When transmitter and
receiver are separated (Section~\ref{sec:model}), timing and carrier-frequency offsets between
their clocks bias the measured delay and Doppler, and must be estimated or canceled before the
bounds of Section~\ref{sec:limits} become attainable.
\item \emph{Hardware impairments.} Phase noise, nonlinear amplifiers, and low-resolution converters
distort both the ambiguity function and the constellation, and must enter the models of
Sections~\ref{sec:radar}--\ref{sec:beamforming}~\cite{lin2026crb,zhang2025distortion}.
\item \emph{Learning-based ISAC.} Data-driven waveform, beamforming, and estimation designs are
promising where accurate models are unavailable, but need interpretability and
guarantees~\cite{demirhan2023integrated,mateos2022end}.
\item \emph{Security and privacy by design.} Embedding confidentiality and sensing integrity into
the waveform and beamformer remains an active frontier~\cite{su2025integrating,welling2024low}.
\end{itemize}

\subsection{Conclusion}
This tutorial has assembled, from first principles, the mathematical foundations of integrated
sensing and communications. Beginning from a unified complex-baseband signal model, we developed the
radar toolkit (ambiguity function, maximum-likelihood estimation, the Fisher information matrix and
Cram\'er--Rao bound, and Neyman--Pearson detection) and the communication toolkit (Shannon and MIMO
capacity, water-filling, and multiuser precoding), then unified them to derive the fundamental limits
of ISAC: the CRB--rate region, the capacity--distortion function, and the deterministic--random
tradeoff. We showed repeatedly that sensing and communication compete over a \emph{shared} design
variable (the transmit covariance $\mat{R}_x$, or the RIS phase $\mat{\Phi}$), and that this sharing
is at once the source of the tradeoff and the origin of the integration gain. Building on the limits,
we treated waveform design, beamforming as explicit optimization, and receiver processing, and we
surveyed the architectures, applications, standardization, and security landscape, closing with a
worked example that ties the layers together. The reader who has followed the derivations now
possesses both the mathematical fluency and the literature map to begin research in ISAC. The central
lesson is a unifying one: sensing and communication are not competing services bolted onto a common
radio, but two projections of a single information-bearing, environment-probing signal; the art
of ISAC is the joint design of that signal.

\section*{Acknowledgment}
The author acknowledges the use of Claude Science, an AI research assistant developed by Anthropic, to assist with drafting, organization, language refinement, and editing of portions of this manuscript. The author has reviewed and verified the technical content, equations, claims, and references, and takes full responsibility for the final manuscript.

\bibliographystyle{IEEEtran}
\bibliography{references}

\end{document}